\documentclass[article,aps,epsf,amsfonts,amssymb,amsmath,nofootinbib]{revtex4}

\usepackage{amsmath,amssymb,amsfonts,amsthm}
\usepackage{graphicx}

\usepackage[english]{babel}

\newcommand{\be}{\begin{equation}}\newcommand{\ee}{\end{equation}}
\newcommand{\bea}{\begin{eqnarray}}\newcommand{\eea}{\end{eqnarray}}
\newcommand{\bsa}{\begin{subeqnarray}}
\newcommand{\esa}{\end{subeqnarray}}
\newcommand{\brr}{\begin{array}}\newcommand{\err}{\end{array}}
\newcommand{\bit}{\begin{itemize}}\newcommand{\eit}{\end{itemize}}
\newcommand{\ben}{\begin{enumerate}}\newcommand{\een}{\end{enumerate}}

\newcommand{\ba}{\begin{array}}
\newcommand{\ea}{\end{array}}

\def\1{{_{1}}}\def\2{{_{2}}}

\def\noHe0{:\;\!\!\;\!\!:H_e(0):\;\!\!\;\!\!:}
\def\noHm0{:\;\!\!\;\!\!:H_\mu(0):\;\!\!\;\!\!:}

\def\1{{_{1}}}\def\2{{_{2}}}

\begin{document}

\title{Transduction of DNA information through water and electromagnetic waves}

\author{Luc Montagnier${}^{a,b}$}
\author{Emilio Del Giudice${}^{c,}$\footnote{deceased 31 January 2014}}
\author{Jamal A\"issa${}^{b}$}
\author{Claude Lavallee${}^{a}$}
\author{Steven Motschwiller${}^{d}$}
\author{Antonio Capolupo${}^{e,f}$}
\author{Albino Polcari${}^{g}$}
\author{Paola Romano${}^{g,h}$}
\author{Alberto Tedeschi${}^{i}$}
\author{Giuseppe Vitiello${}^{e,f}$}
\affiliation{${}^{a}$  World Foundation for AIDS research and Prevention (UNESCO), Paris, France}
\affiliation{${}^{b}$ Nanectis Biotechnologies, S.A. 98 rue Albert Calmette, F78350 Jouy-en-Josas, France}
\affiliation{${}^{c}$ Sezione INFN, I-20122 Milano, Italy (retired) and Centro Studi Eva Reich, Via Colletta, 55, I-20122 Milano, Italy}
\affiliation{${}^{d}$ Chronix Biomedical, GmbH, Goetheallee, 8,  37073 G\"ottingen, Germany}
\affiliation{${}^{e}$ Dipartimento di Fisica E.R.Caianiello
Universit\'a di Salerno, Fisciano (SA) - 84084, Italy}
\affiliation{${}^{f}$ INFN Gruppo collegato di Salerno, Fisciano (SA) - 84084, Italy}
\affiliation{${}^{g}$ Dipartimento di Scienze e Tecnologie, Universit\'a del Sannio, Benevento - 82100, Italy}
\affiliation{${}^{h}$ SPIN-CNR, Universit\'a di Salerno, Fisciano (SA) - 84084, Italy}
\affiliation{${}^{i}$ WHITE Holographic Bioresonance, Via F. Petrarca, 16, I-20123 Milano, Italy}


\begin{abstract}
The experimental conditions by which electromagnetic signals  (EMS) of low frequency  can be emitted by diluted aqueous solutions of some bacterial and viral DNAs are described. That
the recorded EMS and nanostructures induced in water carry the DNA information (sequence)
is shown by retrieval of that same DNA by classical PCR amplification using the TAQ polymerase, including both primers and nucleotides. Moreover, such a transduction process has also been observed in living human cells exposed  to EMS irradiation.  These experiments suggest that coherent long range molecular interaction must be at work in water so to allow the observed features. The quantum field theory analysis of the phenomenon is presented.
\end{abstract}

\maketitle

\section{INTRODUCTION}

This paper is an overview of what we have achieved during the past ten years in this new field of Biology: the role of water memory and electromagnetic waves in biological processes, including pathological conditions.
The reported data is not only of theoretical interest, but
leads to many medical applications.

This work could not have been done and analyzed  without the constant interaction  of biologists and physicists. The quantum field theoretical analysis of the phenomenon points to the crucial role played by coherent molecular dynamics.

\section{ELECTROMAGNETIC SIGNALING OF DNA}

\subsection{ The detection of electromagnetic signals (EMS)}

On 13 July 2005, (the eve of Bastille Day in France), by using a device previously designed  by  the Jacques Benveniste team  to detect electromagnetic signals in water dilutions of biologically active compounds, and with the help of one of his former collaborators, Dr. Jamal A\"issa, two of us (LM, JA) observed for the first time an increase of amplitude and frequency of the recorded electric signals emitted by some high dilutions of filtrates of bacteria (Mycoplasma pirum, then Escherichia coli).
This was the beginning of an extensive investigation on the role and the molecular origin of this new phenomenon (Montagnier,  A\"issa, Ferris, et al., 2009; Montagnier,  A\"issa, Lavallee, et al., 2009; Montagnier,  A\"issa, Del Giudice, et al., 2011).

We soon discovered that DNA was the main source of the initiation of electromagnetic signals in water.  In contrast to the fresh preparation of biological fluids (blood plasma, culture media) which lose their capacity of inducing EMS in water upon freezing, DNA  extraction could be done from frozen material without losing its EMS capacity.

In fact, as we will see below, this property of some bacterial and viral DNA sequences of
emitting EMS is like an indelible tag, and is faithfully transmitted to water structures. The
bacterial species with pathological potential cultured in standard growth media yield DNA
with EMS capacity. However, we noticed that one apathogenic strain of E. coli used for DNA
cloning lacks this capacity as does a probiotic bacterium (Lactobacillus).

The size of DNA fragments emitting EMS ranges between 104 base pair (LTR fragment of HIV) to several kilo-bases (adhesin  gene of Mycoplasma pirum: 1.5 - 3 kbp). Some PCR (Polymerase Chain Reaction) amplicons\footnote{An amplicon is a piece of DNA or RNA that is the source and/or product of natural or artificial amplification or replication events, usually PCR.}, of  400 - 500 bp,  have been found good emitters for the transduction experiments (see below) .
For the capture of EMS in water dilutions, the conditions are very strict, and are the same for the extracted DNA as for the fresh unfrozen samples of plasma or of culture medium:

\noindent {\bf Double filtration}: through 450 nm and then 100 nm Millipore filters, for detection of EMS having a bacterial origin; or 450 nm then 20 nm (Whatman anotop) for EMS of viral origin (only tested for some small virus DNA and HIV DNA).
The usual starting concentration is 2 ng of DNA/1 ml diluted 100 times (10 mls) for filtration.

\noindent {\bf Dilutions}: several decimal  dilutions are made in conic plastic tubes (Eppendorf) usually 0.1 ml/0.9 ml of water under a laminar flow hood. (Fig. 1)
Strong vortex shaking (for 15 seconds) is made at each dilution at room temperature. Water is purchased from commercial firms (usually 5 prime water,  DNase - RNase free from Sigma). Usually, the EMS-emitting dilutions are between $10^{-7}$   to $10^{-13}$ for bacteria, $10^{-6}$  to $10^{-10}$ for small viruses (as HIV1). The first lower dilutions are  apparently ``silent", not emitting detectable signals.

\noindent {\bf Capture of EMS}:
As the EMS are produced by resonance,(Montagnier,  A\"issa, Del Giudice, et al., 2011)  we use either an artificial excitation (7 Hz which was found to be the minimally active frequency) in a closed environment shielded by mu-metal, or in open air exposed to the background ambient noise.
Usually, this background noise predominates in the 50 - 300 Hz range, so that the positive EMS, which are  in the range of  500 - 3000  Hz,   are easily detected.

In the measurement room, cell phones should be turned off (battery removed) as some phones are regulated by low frequency signals.

\subsection{Evidence that EMS emission depends on specific modification of the DNA molecule} The diversity of the DNA sequences emitting EMS does not indicate any clue as to attribute this EMS emission property to specific sequences.

However, we have studied an interesting situation  in the case of HIV-infected patients: here, besides EMS produced by HIV DNA (nanostuctures filtering at 20 nm pores), we detected EMS filtering at 100nm pores  and retained at 20 nm pores produced by DNA of an intracellular bacterium present in red blood cells. Surprisingly, these "bacterial" EMS were found to be produced in part by human DNA sequences integrated in or strongly associated with the bacterial DNA.  The same DNA sequences belonging to the chromosomal genome of the same patient never produced EMS.

Moreover, the same sequence was found present in the red blood cells of some healthy individuals, HIV negative; but in these HIV negative individuals this sequence was found not to emit  signals.  This would indicate that the modification of this DNA resulting in EMS emission occurred only under pathogenic conditions. This modification was maintained in all molecules derived by PCR amplification (amplicon).

As EMS are so far only detected in patients suffering of various chronic diseases, it is tempting to speculate that there is a common biochemical modification of the DNA of infectious bacteria and/or viruses present in such diseases. This modification, which  remains  to be determined,  should be different from a base change (mutation), since  there is no difference in the base sequence of the previously mentioned amplicons in diseased patients,  compared to healthy individuals.

\begin{figure}
\centering \resizebox{6cm}{!}{\includegraphics{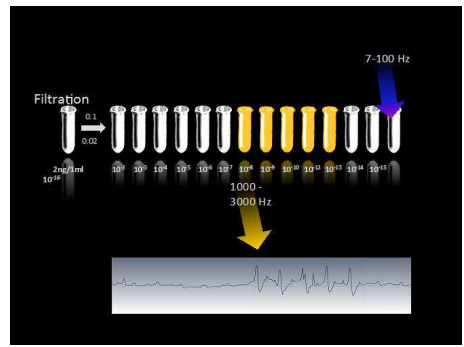}}
\caption{\small \noindent Measuring EMS in water decimal dilutions of DNA. In yellow, dilutions emitting EMS; bottom: recording of EMS in milliseconds}
\end{figure}

\begin{figure}
\centering \resizebox{6cm}{!}{\includegraphics{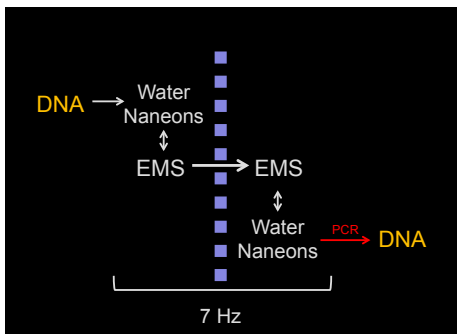}}
\caption{\small \noindent Scheme of DNA transmission through EMS  and water nanostructures (naneons)}
\end{figure}

\begin{figure}
\centering \resizebox{6cm}{!}{\includegraphics{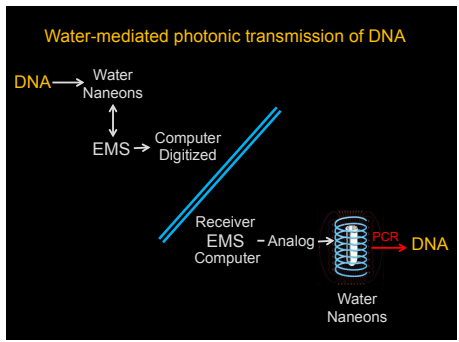}}
\caption{\small \noindent Digital transmission of DNA EMS to a distant laboratory}
\end{figure}

\subsection{ Water nanostructures and EMS do carry DNA information}

Our formerly reported experiments (Fig.~2 and ref. (Montagnier,  A\"issa, Lavallee, et al., 2009)) indicate that the ability of EMS production can be  transmitted from tube 1  containing an emitter DNA dilution to tube 2 of ``naive" water, provided the system is excited overnight by electromagnetic waves of a minimal frequency of 7 Hz. Presumably  tube 1 transmits waves to the water in  tube 2,  which did not originally contain any trace of the DNA at the origin of the signals.

The emission of EMS by the exposed tubes is thus a resonance phenomenon, dependent on external wave input. More importantly, these EMS carry specific information of the initial DNA, as shown by retrieving the DNA by PCR in the recipient tube.

This experiment has been repeated many times in our laboratory, with extraordinary
precautions taken to avoid contamination in the PCR step, and many controls were always
done. Omission of any of the main parameters of the procedure (7 Hz excitation, mu metal
shield, time of exposure to the 7 Hz excitation, any ingredient of the PCR) as well as any
minor detail of the protocol will result in failure of the experiment.

To further make the demonstration unassailable, the  EMS carrying the DNA information were recorded as a  digital file and sent via Internet to a recipient laboratory where work on this DNA or on the bacterium or virus which was the source of that DNA had never been done (Fig.~3). Several labs in Italy and Germany accepted the challenge.

Here, as an example, we show the results obtained at G\"ottingen University using a recorded file (digitized in a lap top computer in our laboratory) of ribosomal 16S DNA from Borrelia burgdorferi (Fig.~4).

In the German laboratory, the electric current resulting from the digital  file communicated by our
lab was converted to analog and was amplified. The current was then connected to a
solenoid. A tube of water was inserted in the solenoid and in this way was submitted to the
induced modulated magnetic field for one hour.
Then the PCR ingredients were introduced in an aliquot of water from the tube, and after 40 PCR cycles of amplification the original DNA was detected, as shown by a specific band in gel electrophoresis of the expected molecular weight.
		
These intriguing results raised several questions :

 1) How a DNA polymerase (the TAQ polymerase of a thermophilic bacterium) can ``read" a genetic code  on water structures?

 2) What about other DNA polymerases of procaryotic and eucaryotic cells? Do they have the same capacity?

Although still at its early stages, the theoretical study of how water structures can store molecular information and transport it by electromagnetic waves gives a crucial role to the coherent molecular dynamics in the formation of water nanostructures (see below and (Montagnier,  A\"issa, Del Giudice, et al., 2011)). We need, however, further theoretical analysis for a complete understanding of the phenomenon, especially because we have  recent evidence that some other DNA polymerases have the same capacity as the TAQ  polymerase to read water messages and can act in living cells.

\begin{figure}
\centering \resizebox{7cm}{!}{\includegraphics{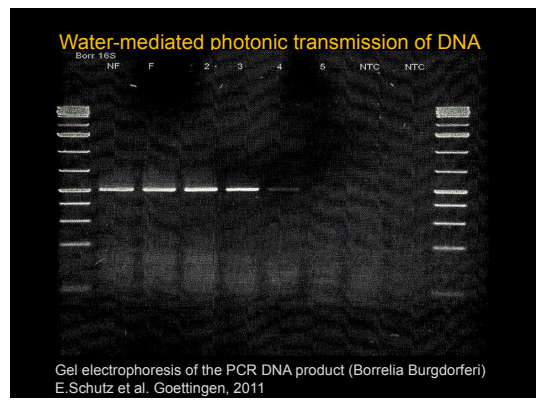}}\label{fig4}
\caption{\small \noindent Gel electrophoresis of Borrelia 16S  DNA amplified by PCR from a water tube having received the DNA specific EMS  (through computer, sound card and amplifier, not shown in this Figure)}
\end{figure}

\subsection{Transduction of DNA in living cells}
The modified DNA transduction system is shown in figure 5.

Instead of magnetizing water in a tube placed inside a solenoid reading the modulated current from the recorded EMS signature, we placed inside the solenoid a flask containing cultured cells; the flask was placed in a  vertical position for cells growing  in suspension, and in a  horizontal position  for cells adhering to the surface of the flask. The voltage (between 2 - 4  volts) applied to the  solenoid was adjusted in order to not generate heat damaging the cultured cells.
This weak intensity was compensated by the duration time of exposure, between 5 to 10 days.
A control flask was always placed outside the solenoid in the same 37 ${}^{\circ} C$ incubator as the exposed flask.

We used several recorded EMS files, including the  16S  Borrelia and the 194 bp  HIV1 LTR amplicon all having been previously shown to be good at transducing their DNA through water.

We tested several immortalized human cell lines derived from leukemias, or cancers: the HL60, originated from a myeloblastic leukemia, the U937, derived from a lung lymphoma, the MCF7, derived from a breast adenocarcinoma. In addition,  we tested  normal cells: the MRC5 diploid fibroblast cell line, derived from the lung of a human embryo, T lymphocytes from a healthy blood donor activated with PHA, and interleukin 2.

Results were striking: All cells of tumor origin synthesized Borrelia 16S DNA after they were exposed for several days to magnetic field modulated by the EMS of Borellia 16S DNA. At the same time, cell growth was inhibited, ending in cell death. DNA was extracted from the dying cells and the Borrelia amplicon was detected by PCR. 

Remarkably the resulting amplicon was found to be EMS emitter, showing that this initial property  was not lost during the complex transmission of DNA information. The  normal differenciated MRC5 cells and the T lymphocytes were not affected in their growth under the same culture conditions and the Borrelia amplicon could not be detected in these cells. The 194 bp HIV LTR amplicon had no effect on the tumor cells.

These preliminary results indicate that the tumoral cell lines so far investigated do possess the enzymatic ability of reading the water nanostructures carrying the DNA information. It remains to be determined whether or not normal embryonic totipotent stem cells have the same ability to read the DNA sequence signals.

\begin{figure}
\centering \resizebox{6cm}{!}{\includegraphics{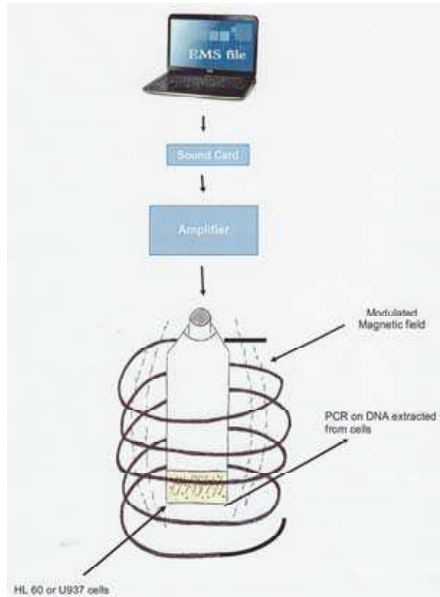}}\label{fig5}
\caption{\small \noindent Scheme of transmission of Borrelia 16S DNA through EMS and naneons into U937  or HL60 cells}
\end{figure}

\section{THEORETICAL ANALYSIS}
In the previous Section we have reported the experimental observation that EMS can be  emitted by diluted aqueous solutions of bacterial and viral DNA under proper conditions. Moreover, it has been observed that duplication of the emitting DNA segment can be obtained by using pure water exposed to  the corresponding DNA EMS and, upon addition of enzymes, primers, etc., submitted to PCR cycles.  Such a transduction process has been observed to occur also in EMS exposed  living cells of tumoral origin. These experimental observations suggest that long range molecular interaction must be at work in water so to allow the observed properties. Indeed, since in the transduction process the high level of sequential ordering among several hundreds of nucleotides entering the transduced DNA chain is obtained, we are clearly in the presence of collective molecular dynamical behavior of water. In quantum field theory (QFT) it is known that the ordering of the elementary components of a system  is achieved as a result of the spontaneous breakdown of symmetry and constitutes the observable manifestation of coherence (Blasone, Jizba and Vitiello, 2011; Fr\"ohlich, 1977; Umezawa, 1993;  Vitiello, 1998;). Ordering is thus not the result of short range forces, but of long range collective coherent correlation.  The classical behavior of the ordered pattern derives from the fact that in coherent states the ratio between the quantum fluctuation $\Delta n $ in the correlation modes and  their condensate number $n$ is $\Delta n/n = 1/{|\alpha|}$ and quantum fluctuations are thus negligible for high $|\alpha|$, which denotes the coherent strength. In the present case, the symmetry which gets broken is the rotational symmetry of the electrical dipoles of the water molecules and correlation modes are the ones associated to the dipole waves (similar to spin waves in ferromagnets) (Del Giudice, Doglia, Milani and Vitiello, 1985, 1986).

We thus conclude that the observed properties of the DNA-water system provide an indication of (may be accounted by) the coherent molecular dynamics. The theoretical analysis based on quantum electrodynamics (QED) shows (Montagnier,  A\"issa, Del Giudice, et al., 2011) that liquid water appears to behave as an active medium able to perform through very low frequency electromagnetic fields (e.m.f.). Short range H-bond and electric dipole-dipole static interactions among liquid water molecules set in as the consequence of the molecule interaction with time-dependent radiative e.m.f. over an extended region called coherence domain (CD) (Del Giudice, Doglia, Milani and Vitiello, 1985, 1986; Del Giudice, Preparata and Vitiello, 1988; Del Giudice and Vitiello, 2006; Bono, Del Giudice, Gamberale and Henry, 2012). Short range H-bond and electric dipole-dipole static interactions are themselves the dynamical effects caused by the most fundamental long range molecular and radiative e.m.f. interaction. This last one is thus responsible for the dynamic origin of short range interactions. This can be better understood by recalling a few points of the discussion presented in (Montagnier,  A\"issa, Del Giudice, et al., 2011).

Above a density threshold and below a critical temperature, an ensemble of molecules
interacting with the e.m.f. undergoes a transition to a dynamical regime characterized
by a minimum energy state where the phase oscillations of the molecules are no longer
uncorrelated. Such a minimum energy state implies a configuration of the system where
all molecules enclosed within the CD oscillate in unison in tune with the e.m.f. trapped
within the CD (phase locking). The linear size of the CD is determined by the wavelength $\lambda$ of the
trapped e.m.f. (typically of the order of 100 nm).
The dynamical mechanism ruling the CD formation is the one of the spontaneous
breakdown of symmetry and it is described in (Del Giudice, Doglia, Milani and Vitiello, 1985, 1986; Del Giudice, Preparata and Vitiello, 1988; Del Giudice, Spinetti and Tedeschi, 2010; Del Giudice and Tedeschi, 2009; Del Giudice and Vitiello, 2006). Its mathematical formulation (Matsumoto, Papastematiou, Umezawa and Vitiello, 1975) is similar to the one of the Anderson-Higgs-Kibble mechanism (Anderson, 1958; Higgs, 1966; Kibble, 1967) which has led to the recent
discovery of the Higgs particle. One important aspect of such a general QFT mechanism is that the transition to the coherent dynamical regime
can be triggered by a vanishingly weak external input. Due to the weakness of the input,
the system does not get ``slaved" by it, but reacts to it according to its own internal
dynamics and, provided that the mentioned conditions of temperature and density are
satisfied, the system sets in a coherent state, whose phase is determined by the phase of
the triggering input (Blasone, Jizba and Vitiello, 2011; Umezawa, 1993). Its coherence strength, however, does not depend on the input
strength\footnote{We stress that a strong input may drive the system  shielding its own
internal dynamics. In such a case, the symmetry is said to be explicitly broken and one has a substantial modification of the original system by inclusion of the strong perturbing agent. However, this is not what
we are interested in in the present case, and in general in Biology, where small perturbing inputs may trigger relevant reaction of the system driven by its own internal dynamics}. In the Appendix we will comment on the question whether coherence is not destroyed by, or theoretically incompatible with the decoherence phenomenon in quantum mechanics (QM). This is an important issue, which shows that the DNA transduction process is indeed a {\it quantum field theory} process; it could not be understood and described in QM, where the decoherence phenomenon does in fact occur (Alfinito, Viglione and Vitiello,  2001). Our framework, however, is the one of QFT.

These features already help us in the understanding of some of the experimental
observations. In particular, it is immediately recognized the observed relevance of
extremely low signal (ELS) in the phenomena under study. The stimulation caused by
the electromagnetic background of very low frequency is indeed observed to be essential
in order for the DNA-water system to emit the EMS. In the experiments, the background
ELS is either produced from natural sources (the Schumann resonances which start at 7.83 Hz (Montagnier,  A\"issa, Del Giudice, et al., 2011; Nickolaenko and Hayakawa, 2002)) or from artificial sources.

The fragment of DNA embedded in the water also acts as a trigger in the surrounding water, causing  the spontaneous formation of CDs, which appear as a self-produced cavity with molecular coherence strength independent from the specific input strength. However,
there is ``phase locking" between the specific DNA molecular structure and the water
molecules. Such a specific feature of the DNA-water coherent coupling accounts for the experimental observations. In the part of the experiment concerning the DNA
transduction, the dynamical phase locking is shared with pure water in a
tube when it is irradiated by the EMS emitted by the aqueous DNA solution system.
For brevity, we omit to report further analysis of the interplay
between the size of the CD, the e.m.f. wavelength and the e.m.f. self-trapping and
frequency inside the CD. For the reader convenience, we report in the Appendix a brief
summary of some other features of CD discussed in (Montagnier,  A\"issa, Del Giudice, et al., 2011).

\begin{figure}
\centering \resizebox{7cm}{!}{\includegraphics{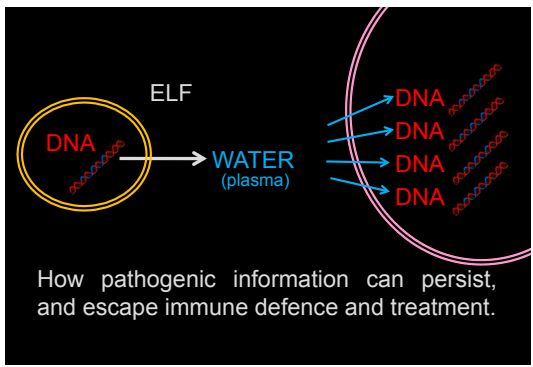}}\label{fig5}
\caption{\small \noindent An hypothetic scheme of the role of EMS in pathogenic DNA amplification.
}
\end{figure}

The question now naturally arises: does the EMS have any specific property related to the coherent dynamical structure discussed above? The question is particularly relevant because the emitted EMS, acting on water molecular dynamics, produces coherent structures such that in PCR processes the DNA transduction occurs with the same nucleotide sequence as the one of
the parent DNA. The answer to the question is provided by observing that the EMS appears to carry not only the specific information of its frequency spectrum, amplitude and phase modulation (the syntactic level),  but it also describes the dynamics out
of which it is generated. In other words, beside the syntactic level of pure information (\`a la Shannon), there is a {\it semantic} content, which manifests itself in the underlying coherent dynamics of the DNA-water system responsible of the polymerization (highly ordered sequence) of hundreds of nucleotides. We refer to such a semantic content as to the ``meaning'' of the EMS. In some recent papers it has been shown that an isomorphism exists between (squeezed) coherent states and self-similar fractal properties (Vitiello, 2009a, 2009b, 2012, 2014). Here, for brevity we do not proceed further with our analysis on this point. Results will be presented elsewhere.

Above we have mentioned the mechanism of phase locking and phase content in water molecular dynamics. Let us close this Section by stressing indeed the crucial role played by the phase in the considered processes.  Due to the relation between phase and electromagnetic potential, non-trivial topological properties, with associated Bohm-Aharonov-like effects, have a non-secondary part in the molecular dynamical properties of water (see e.g.  (Del Giudice, Fuchs and Vitiello, 2010; Del Giudice and Vitiello, 2006) and discussions there reported). Such a remark may turn out to be important when considering charge transfer along the double helix, produced by oxidative agents observed in (Genereux and Barton, 2010). The question if such a current could contribute to the production of EMS may have an affirmative answer since the charge transfer along the DNA produces magnetic field in the surroundings. On the other hand, the oscillations of electric dipoles of the DNA macromolecule may propagate on the DNA in wave form so to contribute to the electromagnetic signal emission. These EMS are considered in (Del Giudice, Doglia, Milani and Vitiello, 1985, 1986) where it is shown that they produce symmetry breakdown in the water in which the dipole chain (DNA or protein chain) is embedded and the mathematical details of the proof are reported.

Finally, concerning the question 1 in Section II.C  (how a DNA polymerase, the TAQ polymerase of a thermophilic bacterium,  can "read" a genetic code on water structures)  the whole dynamical scenario above presented provides the answer to it. It is a complex scenario founded on QFT of coherent systems and is then not surprising that those who are unaware of it could not conceive the positive answer, with all of its complex but clear details, to that question.

\section{CONCLUSION AND PERSPECTIVES}

This 10 year long collaborative work has yielded  some scientific facts and concepts in a new domain of Science at the frontier of Biology and quantum field Physics.

A new property of some DNA molecules has been discovered, that of emitting low frequency electromagnetic waves in water dilutions. These DNAs originate from pathogenic agents or agents  endowed  with pathogenic potential.   It may be not pure coincidence that such EMS are associated with diseases, particularly chronic diseases.

Under  natural conditions  EMS and water nanostructures may play a role of stealthy  elements carrying DNA information  of infectious agents while being undetected by the immune system or being insensitive to conventional therapies (Fig.~6). However, one cannot discard the possibility that  DNA waves can play a role in the physiology of living entities.

Moreover, in the laboratory,  we have shown for the first time that EMS can be re-transcribed  into DNA in living cells. These cells are so far of tumoral origin, opening the way to non-invasive treatments of cancers, assuming that  normal stem cells are not affected, or less affected.

Thus, this new biology that we can call after Jacques Benveniste, Digital Biology, has a very promising future, both at the level of quantum Physics, and in numerous medical applications.

From the point of view of the theoretical understanding of the observed phenomena, the discussion above presented suggests that the dynamical law of coherence acts as a {\it law of form} inducing guided  polymerization processes (controlling morphogenesis): the specific polymerization so obtained is the expression of the semantic content of the EMS mentioned in the previous Section, by us referred to as the signal meaning. The dynamics of coherence appears to play the role of dynamic paradigm ruling the ordering (polymerization) processes through dissipative non-equilibrium dynamics controlled by entropy variations and the consequent appearance of the arrow of time (breakdown of time-reversal symmetry) (Celeghini, Rasetti and Vitiello, 1992; Vitiello, 2012, 2014).

The experiments discussed in this paper suggest that also in the {\it usual} PCR processes the DNA duplication is obtained due to the EMS emitted by the parent DNA in the environment of reciprocal inter-actions with water molecules, enzymes, primers and nucleotides in the solution.

The EMS appears thus to be the carrier of the coherence expressed in the DNA code. One might conjecture (Vitiello, 2014) that modifications are induced in the properties of the EMS resulting in the ``deformation"\footnote{In the jargon of coherent states, the word deformation, or also $q$-deformation,  refers to the technically well defined process of ``squeezing" of the coherent state. For technical details see (Yuen, 1976; Celeghini, De Martino, De Siena, et al. 1995).} of coherence (e.g. such as those, but not only those, induced by the observed bacterial actions; cf. Section II.B). This may play a role in epigenetic modifications, thus revealing  the appearance of ``new meanings"  (in the above mentioned sense) associated to deformed properties of EMS. DNA appears to be the {\it vehicle} through which coherence and its dynamical deformations propagate in living matter  (Vitiello, 2014).

\section*{Acknowledgements}
Partial financial support from MIUR and INFN is acknowledged. We thank Pr. E. Schutz, and Dr. H. Urnovitz, Chronix Biomedical, for allowing transduction experiments in their laboratory.  Mrs. Laila A\"issa is acknowledged for her skilled technician assistance. Mrs. S. McDonnell is acknowledged
for her constant participation in the management of this work.

\appendix

\section{On some properties of water molecular dynamics}
One of the peculiarities of water consists in the fact that water molecules in the CD coherently oscillate between the ground state and an excited state lying at 12.06 eV, just below the 12.60 eV ionization threshold (Bono, Del Giudice, Gamberale and Henry, 2012; Montagnier,  A\"issa, Del Giudice, et al., 2011). The almost free electrons in the CD can be excited by external inputs so to form coherent excitations (vortices), whose entropy is lower than the entropy of the energy supplied by the inputs. Vortices,  due to coherence and their non-trivial topology, are not easily destroyed by small, additional external supplies of energy. On the contrary, additional energetic inputs may add up to form a unique vortex, thus storing in the CDs an amount of energy which may be large enough to activate chemical reactions  among molecules, otherwise below the activation energy threshold. Thus, small energy contributions coming from many high entropy inputs add up to form low  entropy ordered patterns of upgraded high energy  (Del Giudice, Fuchs and Vitiello, 2010; Voeikov and Del Giudice, 2009).

An important remark is that DNA and proteins are polyelectrolytes, and are surrounded by positive counterions. Ions having a cyclotron frequency, $\nu_{c} = q \, B/(2\, \pi\, m)$, where $q$ and $m$ are the electric charge and the mass of the ion, respectively, and $B$ is the magnetic field (Liboff, 1997; Liboff, Smith and McLeod, 1987), may play an important role in obtaining a collective performance of water CDs, a coherence of coherent domains.
The observed dependence of the signal emission on the aqueous dilution may be understood (Montagnier,  A\"issa, Del Giudice, et al., 2011) as follows:  suppose that a low magnetic field (for example a natural or artificially produced background magnetic field) matches the ion cyclotron frequency; suppose it may be then able to extract $n$ ions per CD. Then, due to angular momentum conservation, the plasma of $N$ quasi-free electrons in the CDs starts to counter-rotate with a frequency much higher than the ion cyclotron frequency since electron mass is much smaller than the ion mass. This frequency depends on the number of involved ions, namely on their concentration, which therefore is the only relevant variable. This occurs on all the CDs of the system, whose number is irrelevant for the frequency purpose, in agreement with observations. The magnetic component of EMS is produced by the so induced rotation of the plasma of the quasi-free electrons in the CDs.
As a further effect, a co-resonating field appears in the surroundings of the rotating CDs depending on the ion concentration, i.e. on the DNA solution dilution. It could be at the origin of an extended coherence among CDs. The existence of the observed window of dilutions for the occurrence of the EMS emission could be understood by presuming that the signal produced by the lower dilutions could have a frequency higher than the interval of the values detectable by the used instruments. Higher dilutions, on the contrary, could produce no signal because the ion concentration is decreased below the threshold able to excite the CDs (Montagnier,  A\"issa, Del Giudice, et al., 2011).

We observe that thermal collisions could be in competition with electrodynamic attraction of molecules inside the CD and produce permanent fluxes of molecules between a coherent regime and a non-coherent one, and vice-versa, although the total number of coherent and non-coherent molecules are constant for a given temperature T.
Water is thus not a homogeneous liquid, rather it appears as a two fluid system, with coexisting coherent and non-coherent phases, like in the Landau theory of liquid Helium (Landau, and Lifshitz,1959). We have thus a mixed structure system, consistent also with experimental findings (Taschin, Bartolini, Eramo et. al. 2013), which may appear in observations only when observation time is very short with respect to the time of flickering between the two phases. Near surfaces, the coherent phase may be more stable due to the interaction between water molecules and the surface (Pollack, 2001, 2013). For example, in living matter, water, which is bound to membranes or to biomolecules, could more easily manifest the properties of coherence.

Let us finally consider the question whether the phenomenon of decoherence in quantum mechanics (QM) might contradict our analysis based on the formation and non-vanishing life-time of coherent structures in water. We remark that the {\it belief} that coherence is possible only at very low temperatures is disproved by the fact that coherence is observed in a wide range of temperature, from very low to very high ones: the diamond crystal loses its coherence (it melts) at a temperature of about $+3545 ~{}^{0}C$ in the absence of oxigen; the common kitchen salt $NaCl$ melts at $+804 ~{}^{0}C$; in the iron the coherence of the elementary magnets is lost at $+770 ~{}^{0}C$. In superconductors the critical temperature $T_c$ are much lower, in some compound of niobium they are not higher than $- 252 ~{}^{0}C$ and for some high $T_c$ superconductors it is a little above $-153 ~{}^{0}C$. Also the so called BEC systems (which are mostly condensates of atoms) need very low temperatures and are not so stable. All these systems are macroscopic systems and they are described in quantum field theory in terms of coherent Bose-Einstein condensates. The bosons that condense in a crystal are
called the phonons, i.e. the quanta of the
elastic waves responsible of the ordering in crystals; in the magnets, they are called the magnons, namely the quanta of the spin waves in magnets; in the water, they are called ``dipole wave quanta'' (DWQ), the quanta of the fluctuating molecular dipole waves; and so on. This teaches us that the ordered patterns we observe at a macroscopic scale in these systems are sustained and generated by long range correlations maintained by these waves (Alfinito, Viglione and Vitiello,  2001). One would never be able to construct any of these systems by using short range interaction among the nearest neighbours. Short range interaction, if it is there, is made possible by the long range one which brings ``near'' the components (e.g., making possible the formation of $H$-bonds in water). Decoherence in QM would forbid the existence of crystals, magnets, superconductors, etc.. However, these systems do exist and are observed since they are QFT systems.

\section*{Bibliography}

Alfinito, E., Viglione, R. and Vitiello, G. (2001). The decoherence criterion. Mod. Phys. Lett. B15: 127-135. 

Anderson, P.W.  (1958). Coherent Excited States in the Theory of Superconductivity: Gauge Invariance and the Meissner Effect. Phys. Rev. 110:827-835.

Blasone, M., Jizba, P. and Vitiello, G. (2011).  Quantum field theory and its macroscopic manifestations. London: Imperial College Press.

Bono, I., Del Giudice, E., Gamberale, L. and Henry, M. (2012). Emergence of the Coherent Structure of Liquid Water. Water  4(3): 510-532.
doi:10.3390/w4030510

Celeghini, E., De Martino, S., De Siena, S., Rasetti, M. and Vitiello, G. (1995). Quantum groups, coherent states, squeezing and lattice quantum     mechanics. Annals of Phys.(N.Y.), 241:50-67.

Celeghini, E., Rasetti, M. and Vitiello, G. (1992). Quantum dissipation. Nucl. Phys. B215:156-170.

Del Giudice, E. Doglia, S. Milani, M. and Vitiello, G. (1985). A quantum field theoretical approach to the collective behavior of  biological systems. Nucl. Phys. B251 (FS 13):375-400.

Del Giudice, E. Doglia, S. Milani, M. and Vitiello, G. (1986). Electromagnetic field and spontaneous symmetry breakdown in biological matter. Nucl. Phys. B275 (FS 17):185-199.

Del Giudice, E., Fuchs, E. C. and Vitiello, G. (2010). Collective Molecular Dynamics of a Floating Water Bridge. Water J. 2:69-82.
doi.org/10.14294/WATER.2010.5

Del Giudice, E., Preparata, G. and Vitiello, G. (1988). Water as a free electron laser. Phys. Rev. Lett. 61:1085-1088.

Del Giudice, E. and Tedeschi, A. (2009). Water and the Autocatalysis in Living Matter. Electr. Biol. Med. 28(1):46-52.

Del Giudice, E., Spinetti, P. R. and Tedeschi, A. (2010). Water dynamics at the root of metamorphosis in living organisms. Water. Water J. 2:566-586.

Del Giudice, E. and Vitiello, G. (2006).  Role of the electromagnetic field in the formation of domains in the process of symmetry-breaking phase transitions. Phys. Rev. A 74:022105.

Fr\"ohlich, H.  (1977). Long-range coherence in biological systems. Rivista del Nuovo Cim. 7:399-418.

Genereux, J. C.  and Barton, J. K. (2010). Mechanisms for DNA Charge Transport. Chem Rev. 110(3):1642-1662.

Higgs, P. W. (1966). Spontaneous symmetry breakdown without massless bosons. Phys. Rev. 145:1156.

Kibble, T.W. B. (1967). Symmetry breaking in non-Abelian gauge theories. Phys. Rev. 155:1554

Landau, L.D. and Lifshitz, E. M. (1959). Fluids Mechanics. Reading, Mass.: Addison-Wesley, Ch. XVI.

Liboff, A. R. (1997). Electric-field ion cyclotron resonance. Bioelectromagnetics 18(1):85-7.

Liboff, A. R. Smith, S. D.,  McLeod, B. R. (1987). Experimental Evidence for Ion Cyclotron Resonance Mediation of Membrane Transport. In Mechanistic Approaches to Interactions of Electric and Electromagnetic Fields with Living Systems, Editors: Blank, M., Findl, E.. New York:Springer-Verlag,
pp 109-132.

Matsumoto, H., Papastamatiou, L., Umezawa, H. and Vitiello, G. (1975). Dynamical rearrangement of Symmetry in the Anderson-Higgs-Kibble Mechanism. Nucl. Phys. B97:61-89.

Montagnier, L., A\"issa, J., Ferris, S., Montagnier, J-L., Lavallee, C. (2009). Electromagnetic Signals Are Produced by Aqueous Nanostructures Derived from Bacterial DNA Sequences. Interdiscip. Sci. Comput. Life Sci. 1:81-90.

Montagnier, L., A\"issa, J., Lavallee, C., Mireille Mbamy, M., Varon, J., Chenal, H. (2009). Electromagnetic detection of HIV DNA in the blood of AIDS patients treated by
antiretroviral therapy. Interdiscip Sci. Comput. Life Sci. 1:245-253.

Montagnier, L., A\"issa, J., Del Giudice, E. Lavallee, C., Tedeschi, A., Vitiello, G. (2011). DNA waves and water. Journal of Physics: Conference Series 306:012007.

Nickolaenko, A. P. and Hayakawa M. (2002). Resonances in the Earth-ionosphere Cavity. Dordrecht-Boston-London: Kluwer Academic Publishers.

Pollack, G.H. (2001). Cells, Gels and the Engines of Life. Seattle, WA: Ebner and Sons.

Pollack, G.H. (2013). The fourth phase of water: Beyond Solid, Liquid, and Vapor. Seattle, WA: Ebner and Sons.

Taschin, A., Bartolini, P.,  Eramo, R. et. al. (2013). Evidence of two distinct local structures of water from ambient to supercooled conditions. Nature Comm. 4:2041-2045.\\
doi:10.1038/ncomms3041

Umezawa, H.  (1993). Advanced Field theory: micro, macro and thermal concepts. New York: American Institute of Physics.

Vitiello, G.  (1998). Structure and function. An open letter to Patricia Churchland. In Hameroff, S.R. Kaszniak, A. W. and Scott, A.C. Eds.. Toward a science of consciousness II. The second Tucson Discussions and debates. Cambridge: MIT Press. p. 231-236.

Vitiello, G. (2009a). Coherent states, Fractals and brain waves. New Mathematics and Natural
Computation 5: 245-264.

Vitiello, G. (2009b). Fractals and the Fock-Bargmann representation of coherent states. In P. Bruza, D. Sofge, et al. Eds., Quantum Interaction. Lecture Notes in Artificial Intelligence, Edited by  R.Goebel, J. Siekmann, W.Wahlster. Berlin: Springer-Verlag, p. 6-16.

Vitiello, G. (2012). Fractals, coherent states and self-similarity induced noncommutative geometry. Phys. Lett. A 376:2527-2532.

Vitiello, G. (2014). On the Isomorphism between Dissipative Systems, Fractal Self-Similarity and Electrodynamics. Toward an Integrated Vision of Nature. Systems 2:203-216.
doi:10.3390/systems2020203

Voeikov, V. L. and Del Giudice, E. (2009).  Water respiration: the base of the living state. Water J. 1:52-75.

Yuen, H. P. (1976). Two-photon coherent states of the radiation field. Phys. Rev. A 13:2226.

\end{document}